\documentclass[aps,prev,onecolumn,preprintnumbers,aps,
showkeys,floatfix,nofootinbib,brazil]{revtex4-1}

\usepackage[utf8]{inputenc}
\usepackage{lmodern}			
\usepackage[T1]{fontenc}		
\usepackage[brazil]{babel}	
\usepackage{graphicx}
\usepackage{color}
\usepackage{dcolumn}
\usepackage{bm}
\usepackage{epsf}
\usepackage{mathtools}
\usepackage{mathptmx}
\usepackage{yhmath}
\usepackage{amsmath}
\graphicspath{{figArticle/}}
\usepackage{float}

\usepackage{multirow}

\begin{document}
\title{The growth factor parametrization versus numerical solutions in flat and non-flat dark energy models.}

\author{A. M. Velasquez-Toribio}
\email{alan.toribio@ufes.br}
\affiliation{Nucleo Cosmo-ufes \& Departamento de F\'{\i}sica, Universidade Federal do Espirito Santo,  29075-910 Vit\'{\o}ria - ES, Brasil}

\author{Júlio C. Fabris}
\email{julio.fabris@cosmo-ufes.org}
\affiliation{Nucleo Cosmo-ufes \& Departamento de F\'{\i}sica, Universidade Federal do Espirito Santo,  29075-910 Vit\'{\o}ria - ES, Brasil} \affiliation{National  Research  Nuclear  University  MEPhI,  Kashirskoe  sh.   31,  Moscow  115409, Russia}

\date{\today}

\begin{abstract}
In the present investigation we use observational data of $ f \sigma_ {8} $ to determine observational constraints in the plane 
$(\Omega_{m0},\sigma_{8})$ using two different methods: the growth factor parametrization and the numerical solutions method for density contrast, $\delta_{m}$. We verified the correspondence between both methods for three models of accelerated expansion: the $\Lambda CDM$ model, the $ w_{0}w_{a} CDM$ model and the running cosmological constant $RCC$ model. In all case we consider also curvature as free parameter. 
The study of this correspondence is important because the growth factor parametrization method is frequently used to discriminate between competitive models. Our results we allow us to determine that there is a good correspondence between the observational constrains using both methods. 
We also test the power of the $ f\sigma_ {8} $ data to constraints the curvature parameter within the $ \Lambda CDM $ model. For this we use a non-parametric reconstruction using Gaussian processes. Our results show that the $ f\sigma_ {8}$ data with the current precision level does not allow to distinguish between a flat and non-flat universe.
\end{abstract}
\maketitle

\section{Introduction}

The accelerated expansion of the universe is one of the biggest problems in current cosmology, 
since there is no coherent explanation for this accelerated expansion.
Initially, it was associated with a cosmological constant or vacuum energy and 
subsequently models with scalar fields (also known as quintessence models) were evoked. 
Other possibilities include modified gravitation, extra dimensions, and so on. 
For a recent review, see references \cite{pebblesaa, clifton, carroll}.

The main evidence of accelerated expansion is based on background observations, basically on cosmological distance measurements using Supernovas Ia \cite{perlmutter, riess}. However, data on large-scale structure formation are essential to characterize accelerated expansion. The most remarkable example is the cosmic background radiation. Thus, for example, recent measurements from the PLANCK satellite have allowed to measure the values of cosmological parameters with unprecedented precision. Currently we can say that the observational evidence of accelerated expansion is robust using various independent and complementary observational data \cite{esenstein, tanabashi, scolnic, planck, des}.

Additionally, large surveys of galaxies are essential to discriminate between competitive cosmological models that characterize the accelerated expansion of the Universe. A fundamental tool to distinguish between dark energy models or models including new physics is the linear growth factor. Observationally this factor can be derived from the study of the perturbations of the galaxy density $ \delta_ {g} $, which is related to the perturbation of the matter through the bias parameter: $ \delta_ {g} = b \delta_ {m} $, being that the bias, $ b $, can vary between the values $b \in (1,3)$. Therefore, it is difficult to use the linear growth factor, $f = \frac {d\ln\delta_{m}} {d\ln a}$,  as a cosmological test to constrain parameters. In this sense, a more feasible observable turns out to be the product $ f \sigma_ {8} $ \cite{percivall} where $ \sigma_ {8} $ is the variance of the linear matter perturbations 
within spheres of radius $R =8 h^{-1} Mpc$. This observable can be determined using RSD (Redshift Space Distortion) observations, as well as, weak lensing measurements.

For a given cosmological model the observable $ f \sigma_ {8} $ can be theoretically determined using the linear perturbation theory. However, there is an alternative approach which involves introducing a parametrization for linear growth factor. This parametrization was initially proposed and developed by Peebles \cite{pebblesa, pebblesb, peebbles} considering that the linear growth factor must be directly proportional to the matter parameter, which can be adjusted for a given cosmological model.
 
For example, in the case of the flat $ \Lambda CDM $ model the parametrization is $ f = \Omega_{m} ^ {\gamma} $, where $ \gamma $ is a constant around $ \gamma = 6/11 $. In the literature this parametrization has been intensively used to study the growth of the structures. Thus, Lightman and Schechter \cite {lightman} studied the linear growth factor to determine the peculiar velocity in the case of a universe dominated by matter and a perturbation of spherical density. Lahav et al., \cite{lahav} considered the linear growth factor in a Universe with matter plus a cosmological constant and determined an approximate form given by: $ f (z=0) = \Omega_{m0} ^ {0.60} + (1+ \frac{\Omega_{m0}}{2}) \frac{1 }{70}\lambda_{0} $, where $\lambda_{0}=\frac{\Lambda}{3H_{0}^{2}}$. Later, this parametrization was reintroduced into the paradigm of the accelerated expansion of the Universe by Wang and Steinhartd \cite {wang}, and this idea was expanded by Linder \cite{linder2002,linder2004,linder5}, among others. Recently, this approach has been widely used to discriminate between modifying gravity models versus dark energy models, see references \cite{nesserisss, nesseris, gravitation, huterer, belloso, skara, perez}.

In general, to study dynamical dark energy models it is necessary to introduce the called growth index, $\gamma(z)$, which can depend on the redshift. In this sense, studies have been carried out on the global mathematical properties of the growth index, which allows us to study the general characteristics of the dynamics of cosmological models, see recent references \cite{calderon,calderonn}.

Therefore, an important question is to investigate the statistical compatibility between the observational constraints determined using the growth factor parametrization and the constraints obtained using numerical solutions. This question is essential to consider the growth factor as a useful tool to discriminate between competitive models. Thus, in this article we focus explicitly on this question. We study three cosmological models: the $ \Lambda CDM $ model, $ w_{0} w_{a} \Lambda CDM $ model and the running cosmological constant,$ RCC $, model, in all cases we also consider the non-flat models. Additionally, we study the power of the $ f \sigma_{8} $ data to constraint the curvature parameter. For this we use the non-parametric method of Gaussian processes.

Our paper is organized as follows. In Section II, we briefly presented the cosmological models studied and the reconstruction non-parametric. In Section III is devoted to briefly consider observational data. In Section IV, we present our results and in Section V our conclusions.

\section{Dark energy models}

\subsection{Flat and non-flat $\Lambda CDM$}
The cosmological standard model is the $ \Lambda CDM $ model which fits a large amount of observational data very well,
however, certain tensions have arisen in the statistical correspondence of cosmological parameters. For example, the $H_{0}$ tension: local measurements of the Hubble parameter have a tension of at least $4 \sigma$ with measurements of $H_{0}$ using data from the Planck Collaboration \cite{planck}. Also, some researchers have determined a certain curvature tension \cite{handley}, 
this is, many observational data are statistically better fit for closed curvature models, including Planck lensing data 
\cite{ooba}-\cite{amvt}. In the present work we are interested in investigating the implications of introducing curvature in the study of linear growth factor. Therefore, we consider the non-flat $\Lambda CDM $ model, where the Hubble parameter is given by,
\begin{eqnarray}
H = H_{0}\sqrt{\Omega_{m0}(1+z)^{3}+ \Omega_{k0}(1+z)^{2} + \Omega_{\Lambda 0}},
\end{eqnarray}
where we use the definitions: 
\begin{equation}
\begin{array}{lll}
\Omega_{m0} = \frac{8\pi G \rho_{m0}}{3H_{0}^{2}}, & \Omega_{\Lambda 0} = \frac{\Lambda}{3H_{0}^{2}} &  {\rm and}~\Omega_{k0} = \frac{-k}{a^{2}H_{0}^{2}}.
\end{array}
\end{equation}
where $k$ is the spatial curvature which can be $k = + 1$ for a closed universe, $k=0$ for a flat Universe and $k = -1$ for an open universe. Additionally we have the restriction,
\begin{eqnarray}
\Omega_{m0}+\Omega_{\Lambda 0}+\Omega_{k0} =1.
\end{eqnarray}
The $ H (z) $ function allows to fully characterize the cosmological model at the background level, but to study the growth of structures it is necessary to introduce deviations from the background. To do this we consider the theory of cosmological perturbations initiated by Lifshitz \cite{lifshitz} which is, in the linear regime, a well established consistent theory \cite{perturbation,perturbationn,efstathiou}.

\subsubsection{Numerical Solution}
Considering the theory of cosmological perturbations the evolution of matter fluctuations, $\delta_{m}= \frac{\delta \rho_{m}}{\rho_{m}}$, 
is governed by the equation,
\begin{eqnarray}
\ddot{\delta}_{m}(t) + 2H\dot{\delta}_{m}(t) - 4\pi G \rho_{m} \delta_{m}(t) &=& 0,
\end{eqnarray}
where the derivative is with respect to cosmic time. However, for our calculations it is more convenient to rewrite the previous equation in function of the redshift obtaining the equation,
\begin{eqnarray}
\delta''_{m}(z) + \left(\frac{H'}{H} - \frac{1}{1+z}\right)\delta'_{m}(z) - \frac{3}{2}(1+z)\frac{H_{0}^{2}}{H^{2}}\Omega_{m0}\delta_{m}(z) &=&0
\end{eqnarray}
This equation has been extensively studied and in the case of a flat $\Lambda CDM$ model there are analytical solutions, see references 
\cite{eheath} - \cite{eisenstein},
\begin{eqnarray}
\delta_{m}(a) & = & a {}_2F_{1}\left(-\frac{1}{3w},\frac{1}{2}-\frac{1}{2w},1-\frac{5}{6w},a^{-3w}(1-\frac{1}{\Omega_{m}})\right),
\end{eqnarray}
where the ${}_2F_1$ is a hypergeometric function. In the case of a non-flat Universe there are no analytical solutions, except for some particular and approximate cases such as those published by Hamilton \cite{hamilton}. In the present paper we obtain theoretical solutions of the observable $ f \sigma_{8} $ including curvature. This is done by numerically solving the equation for the density contrast $ \delta_{m} $.

On the other hand, the observable $\sigma_{8}(z)$ is the redshift-dependent rms fluctuations of the linear density field at 
$R=8h^{-1}$ Mpc and is given by
\begin{equation}
\sigma_{8}(z) = \sigma_{8}\frac{\delta_{m}(z)}{\delta_{m}(0)}
\end{equation}
where $\sigma_{8}$ is the currently value. Therefore, with some approximations on the linear scale we can derive the observable $ f\sigma_{8} $ from the solution of the equation (4), but first we define
\begin{eqnarray}
f(a) \equiv \frac{d \ln \delta_{m}}{d \ln a} = -(1+z)\frac{d \ln \delta_{m}}{d z}
\end{eqnarray}
and using these definitions, we can write, 
\begin{equation}
f(z)\sigma_{8}(z) = -\sigma_{8}(1+z)\frac{\delta'_{m}(z)}{\delta_{m}(0)}
\end{equation}
This observable can be used to constraints cosmological parameters using observational data determined from (Redshift-space distortions) RSD measurements.
\subsubsection{Growth factor parametrization}
As mentioned, the growth index is a way to simplify the calculations and is strongly based on theoretical considerations. We can explicitly define the linear growth rate $ f $ in the form,
\begin{equation}
f(z) = \frac{d \ln \delta_{m}}{d \ln a} \approx \Omega_{m}^{\gamma}(z).
\end{equation}
The initial motivation for this parametrization is the paper of Peebles \cite{pebblesa}, 
where the author considers the case of a universe dominated by matter
and shows that the growing solution of the equation (4) is directly proportional to $ \Omega_{m} $. 
It is also possible to motivate this parametrization for quintessence models by following the \cite{wang} 
reference, where the equation (4) can be written as a function of $ f $. The authors determine the gowth index as $ \gamma = \frac {6}{11} + \frac {3}{200} (1- \Omega_{m})+ O(2) $. Therefore, a good approximation for the flat $ \Lambda CDM $ model  is $ \gamma \approx \frac {6} {11} $.
Numerous investigations have used this parametrization 
to study cosmological models, in particular
see the references \cite{calderon, calderonn}. In the present work we extend this parametrization to include curvature.
Therefore, based in this equation we can obtain,
\begin{equation}
D(z) \equiv \frac{\delta_{m}(z)}{\delta_{m}(0)}= \exp\left[-\int_{0}^{z}\frac{\Omega_{m}^{\gamma}}{(1+x)} dx\right].
\end{equation}
We consider $D(z)$ as normalized to unity at the present time. Therefore, using this parametrization the observable $f \sigma_{8}$ is given by
\begin{equation}
f\sigma_{8}(z) = \sigma_{8} D(z) \Omega_{m}^{\gamma}
\end{equation}
where we have set $\sigma_{8}(z)=\sigma_{8} D(z)$. In the case of $ \Lambda CDM $ we consider the growth index, $\gamma$, as constant.

\subsection{Dynamical Dark Energy: CPL-Parametrization}
One way to relax the cosmological constant is to introduce a parametrization that varies over time. A fairly popular parametrization that allows the inclusion of a wide family of cosmological models and that somehow retains a certain simplicity is the called CPL-parametrization given by the form \cite{linder2002, linder2004},
\begin{equation}
w(z) = w_{0}+w_{a}\frac{z}{1+z},
\end{equation}
where $w_{0} $ represents the cosmological constant and note that $(\frac{dw(z)}{dz})_{z=0}=w_{a}$ one might consider this quantity a natural measure of time variation. The CPL parametrization describes fairly gradual evolution
from a value of $w = w_{0} + w_{a}$ at early times to a present-day value of $w = w_{0}$.
Thus we can write the Hubble parameter,

\begin{equation}
H(z) = H_{0}\sqrt{\Omega_{m0}(1+z)^{3} + \Omega_{k0}(1+z)^{2} + (1-\Omega_{m0}-\Omega_{k0})(1+z)^{3(1+w_{0}+w_{a})}e^{\frac{-3w_{a}z}{1+z}}},
\end{equation}
where we have included the curvature parameter. This model has been extensively used as a two parameter parametrization paradigm. In the reference \cite{padmanbhan, scherrer} it was shown that thawing quintessence models with a nearly flat potential all converges toward the behavior given by $-1.5(1 + w_{0}) $. Therefore, this parametrization allows extrapolating results for quintessence-type models.

\subsubsection{Numerical Solution}
In this case the cosmological perturbation theory allows us to write an equation for the fluctuations of matter, $\delta_{m}$ analogous to the equation (5), however considering the Hubble parameter given by the previous equation (15). The calculation process is similar to that developed for the $ \Lambda CDM $ model.

\subsubsection{Growth factor parametrization}
The parametrization of the linear growth rate for this case is given by the expression,
\begin{equation}
f=\Omega_{m}(z)^{\gamma(z)}
\end{equation}
where we now consider $ \gamma $ as a function of the redshift $ \gamma (z) $. This function is introduced to quantify the effects of a dynamic dark energy model. Several functions have been introduced as ansatz for the function $ \gamma (z) $, in this paper we are going to use the following form,

\begin{equation}
\gamma(z)= \gamma_{0} + \gamma_{a} \frac{z}{1+z}.
\end{equation}
This functional form for $\gamma$ is well-behaved for late redshift values and therefore can be suitably used for $ f \sigma_ {8} $ data that includes data up to approximately $ z \approx 2.0 $.

\subsection{Running Cosmological Model}
This cosmological model is based on the results of the renormalization group applied to cosmology. Specifically, a quadratic model for the cosmological constant was presented in \cite{shapiro2000}-\cite{shapiro2004} called the running cosmological constant ($RCC$) model. Furthermore, this model was extended for the case of a gravitational logarithmic coupling \cite{shapiro2005}. In the present work we study the quadratic model for the cosmological constant. In this model, the energy density of the vacuum 
$ \rho_{\Lambda}(z) $ can be given as a quadratic function of the rate of expansion,
\begin{equation}
\rho_{\Lambda} = \rho_{\Lambda_{0}} + \frac{3 \nu M_{pl}^{2}}{8 \pi} (H^{2}-H^{2}_{0}),
\end{equation}
where the $\nu$ parameter is given by,
\begin{equation}
\nu = \frac{\sigma M^{2}}{12 \pi M_{p}^{2}},
\end{equation}
the parameter $M$ is an effective mass parameter representing the average mass of the heavy particles in the grand unified theory $(GUT)$
near the Planck scale, after taking into account their multiplicities.
The coefficient $\sigma$ can be positive or negative, the sign depends on whether bosons $(\sigma=+1)$ or fermions $(\sigma=-1)$ dominate in the loop contribution, this is, it depends on whether fermions or bosons dominate at the highest energies.
In this framework, the energy density in the $RCC$ model is,
\begin{equation}
\frac{\rho_{\Lambda}}{d \ln H}  =  \frac{\sigma  H^{2} M^{2}}{16\pi^{2}},
\end{equation}
which was proposed based on the assumption that the renormalization group scale $\mu$ is identified with
$H(z)$. This scale was originally proposed in \cite{shapiro2000,shapiro2002} 
and it is based on the scale dependency in the renormalization group framework.
Thus using the Friedmann equation and the conservation law we can to determine the Hubble parameter $H(z)$ as function of the redshift,
\begin{equation}
\frac{H^{2}}{H^{2}_{0}} = 1 + (\Omega_{m0}-\frac{2 \nu \Omega_{k0}}{1-3\nu})(\frac{(1+z)^{3-3\nu}-1}{1-\nu}) + \frac{\Omega_{k0}(z^{2}+2z)}{1-3\nu}
\end{equation}

\subsubsection{Numerical Solution}
The linear perturbations for the RCC model have been studied in various papers, for example, see \cite{grande,bsola} and in the Newtonian gauge, see \cite{velasquez}. Also more recently using various observational data \cite{sola}. Therefore, we can write for the perturbations of matter,
\begin{equation}
\ddot{\delta_{m}} + (2H+Q)\dot{\delta_{m}} - (4 \pi G \rho_{m}-2HQ-\dot{Q})\delta_{m}=0
\end{equation}
where point indicates derivative with respect to cosmic time. The factor $Q$ represents the variable cosmological constant and is defined as,
\begin{equation}
Q=\frac{\dot{\rho}_{\Lambda}}{\rho_{m}}.
\end{equation}
For our case it is more convenient to rewrite the previous equation in function of the redshift as,
\begin{eqnarray}
\frac{d^{2}\delta_{m}}{dz^{2}} + \left[\frac{d \ln H}{dz}-\frac{1}{(1+z)}\left(1+\frac{Q}{H}\right)\right]\frac{d\delta_{m}}{dz} = \left(\frac{3}{2}\Omega_{m}-\frac{2Q}{H}
+\frac{(1+z)}{H}\frac{dQ}{dz}\right)\frac{\delta_{m}}{(1+z)^{2}}.
\end{eqnarray}
If we consider the condition that $Q = 0$, then the above equation reduces to equation (5), as expected.

\subsubsection{Growth factor parametrization}
In this case the same prescription of the previous cases is also followed so we can write the growth factor in accordance with the reference \cite{bsola},
\begin{equation}
f = \frac{d \ln \delta_{m}}{d \ln a} \approx \tilde{\Omega}_{m}^{\gamma(a)} = \frac{\Omega_{m}(a)^{\gamma(a)}}{1-\nu},
\end{equation}
where we use the same form for the growth index as the previous case given by the equation (15). The function $ \Omega_{m0}(a) $ is given by
\begin{eqnarray}
\Omega_{m}(a) = \frac{\Omega_{m0} a^{-3(1-\nu)}}{H^{2}(a)/H^{2}_{0}}
\end{eqnarray}
where the $H^{2}(a)/H^{2}_{0}$ is given by the equation (19) and if $\nu=0$ again reduces to the case of $\Lambda CDM$ model.
In this case we consider the parameter $\tilde{\Omega}_{m}$  instead of $\Omega_{m}$, since if we consider the regime at high redshift $z>>1$, the matter parameter is $\Omega_{m} \approx (1-\nu)$. In this way for our parametrization given by the equation (24) we obtain
at large redshift $z>>1$  the normalized value of approximately $f \approx 1$.

\subsection{Curvature and $f\sigma_{8}$ data in the non-flat $\Lambda CDM$}
We investigated the power of the data $f \sigma_{8}$ to constrain the curvature parameter. For this we reconstruct the observable 
$ f\sigma_{8} $ directly from observational data using the non-parameter method of Gaussian processes \cite{gaussian}.
We compare this reconstruction with the best fits for the flat $ \Lambda CDM $ case, as well as, for the non-flat model. In principle this allows us to observe the effect of the curvature parameter. To carry out this reconstruction we use the public package Gapp \cite{seikel}. 

\section{Observational Data}
In the present investigation we use as observational data the $ f \sigma_ {8} $ data, which are independent of the bias and may be obtained using the redshift space distortion (RSD) technique. The data used are the data compiled in the reference \cite{kasantzidis}
and consists of $63$ datapoints. In this case the chi-squared is given by the expression,
\begin{eqnarray}
\chi^{2}_{f\sigma_{8}} = V^{i}_{f\sigma_{8}}C^{-1}_{ij} V^{j}_{f\sigma_{8}}
\end{eqnarray}
where the $C^{-1}_{ij}$ is the inverse covariance matrix. We assume that it is a matrix diagonal except for WiggleZ data subset, in this case we have,
\begin{eqnarray}
C_{ij}^{WigglesZ} = 
\begin{bmatrix} 
6.400 & 2.570 & 0.000 \\
2.570 & 3.969 & 2.540  \\
0.000 & 2.540 & 5.184  \\
\end{bmatrix} ,
\end{eqnarray}
Also the vector $V^{i}$ is defined as:
\begin{eqnarray}
V^{i} &=& f\sigma_{8}^{obs} - \frac{f \sigma_{8}^{the}}{q(z_{i},\Omega_{m0},\Omega_{m0}^{fiducial})},
\end{eqnarray}
where $f \sigma_{8}^{the}$ is the theoretical prediction of the $f \sigma_{8}$ observable for each cosmological model.
The $q$ represents the fiducial correction factor introduced in the reference \cite{kasantzidis}.
If an incorrect cosmology is adopted when converting redshift to distance, then the apparent spatial distribution of galaxies will be distorted, therefore, it is necessary to introduce a correction factor, $q$, which can be defined as \cite{kasantzidis},
\begin{eqnarray}
q=\frac{H(z) d_{A}}{H^{fid}(z) d_{A}^{fid}},
\end{eqnarray}
where $ d_{A} $ is the angular distance. The numerator corresponds to the best fit of the $\Omega_{m0}$ parameter in the studied cosmology and the denominator corresponds to the fiducial cosmology of each survey. 
We minimize the chi-squared to obtain the observational constraints on the cosmological parameters.

 \begin{figure*}[htbp] 
 	\centering
	\includegraphics[scale=0.400]{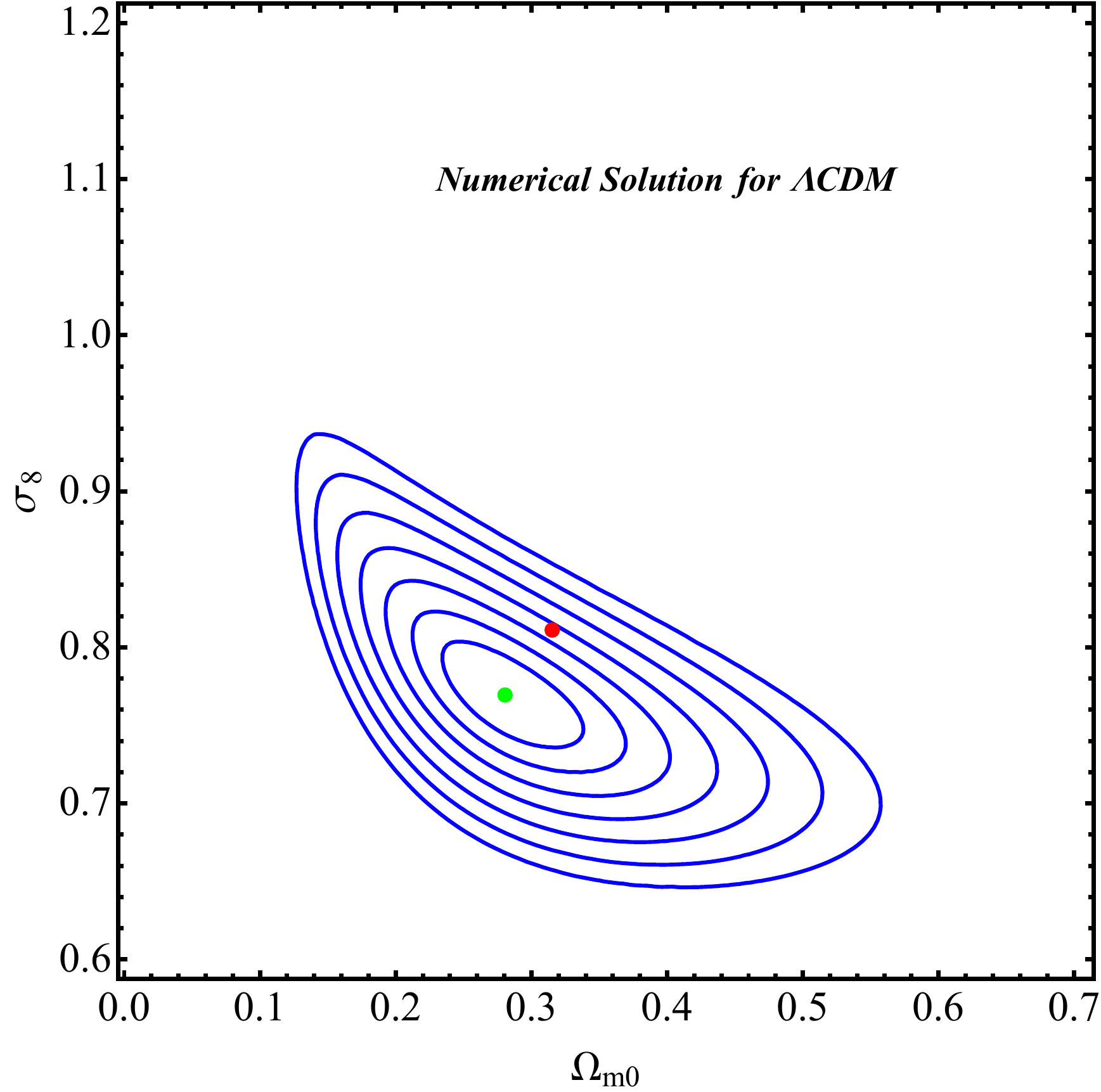}
		\includegraphics[scale=0.400]{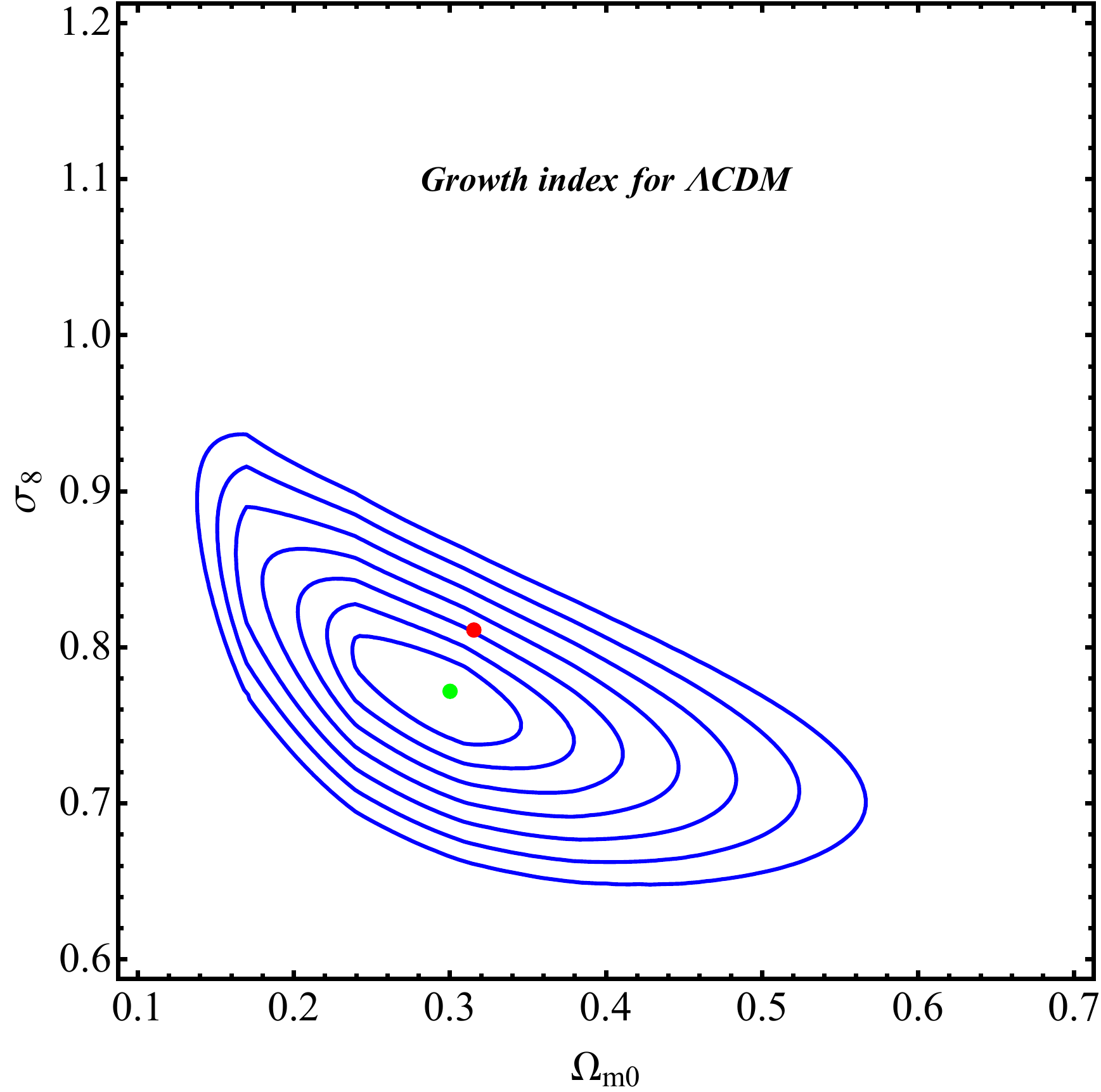}
 		\includegraphics[scale=0.400]{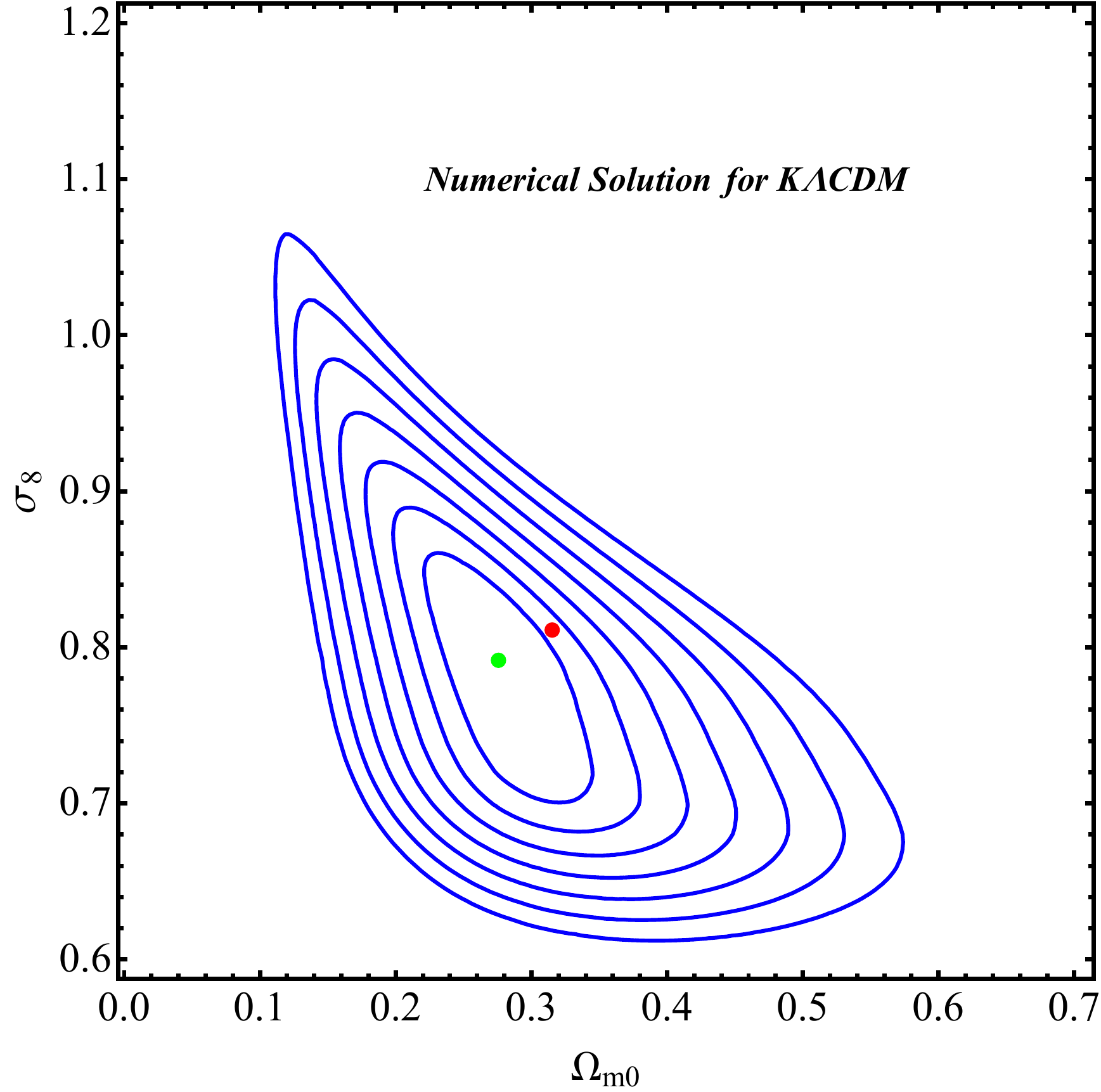}
		\includegraphics[scale=0.400]{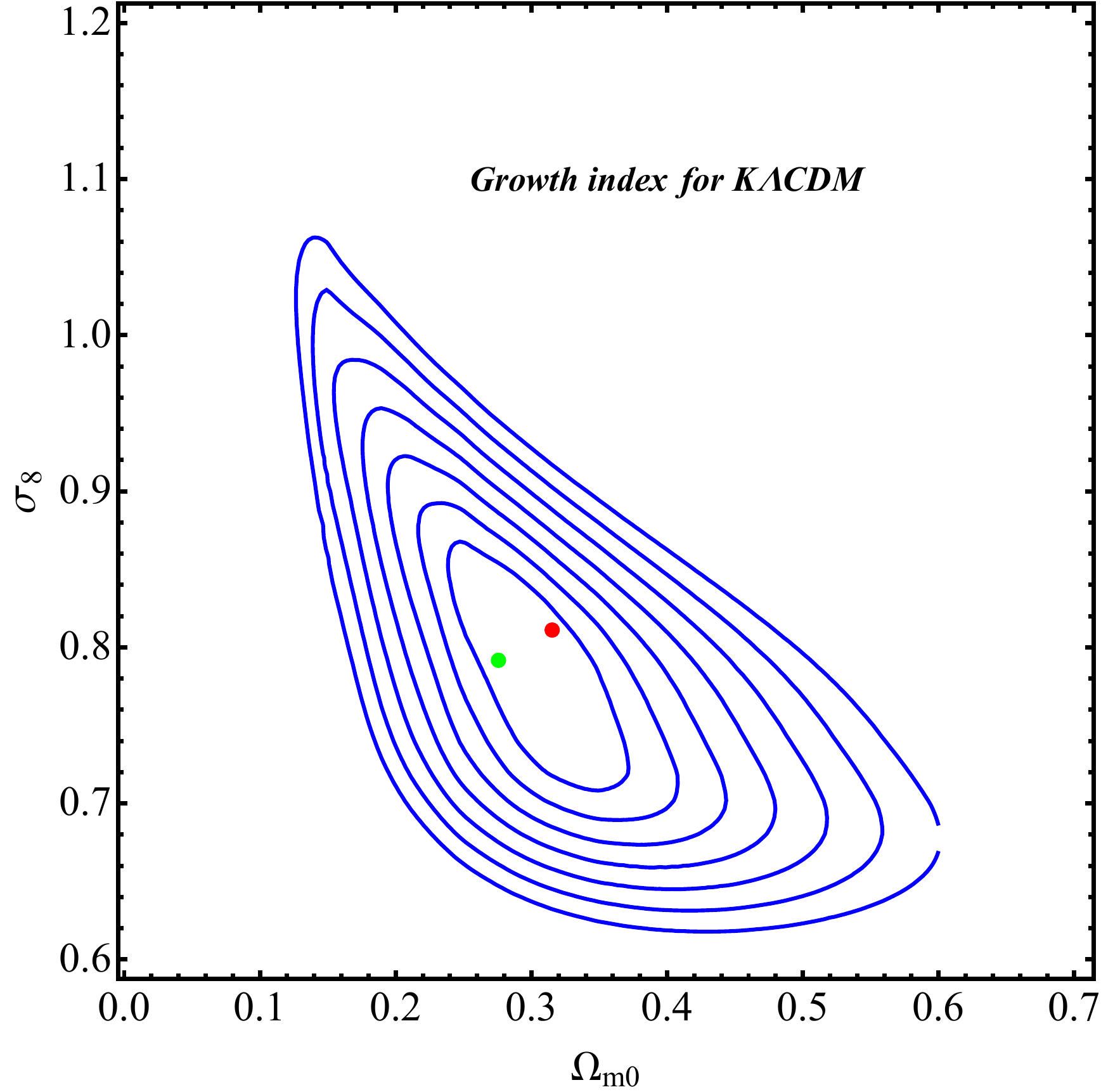}
 	\caption{In the top, we can see observational constraints on the flat $\Lambda CDM$ model.
 	In the bottom, we see the observational constraints on the non-flat $\Lambda CDM$ model.
	The $\Omega_{k0}$ parameter was marginalized in the range of
$-0.1<\Omega_{k0}<0.1$. The green point is the best fit and red point is the Planck result.}
 	\label{fig:epsilon_01_03}
 \end{figure*}

\begin{figure*}[htbp] 
 	\centering
	\includegraphics[scale=0.400]{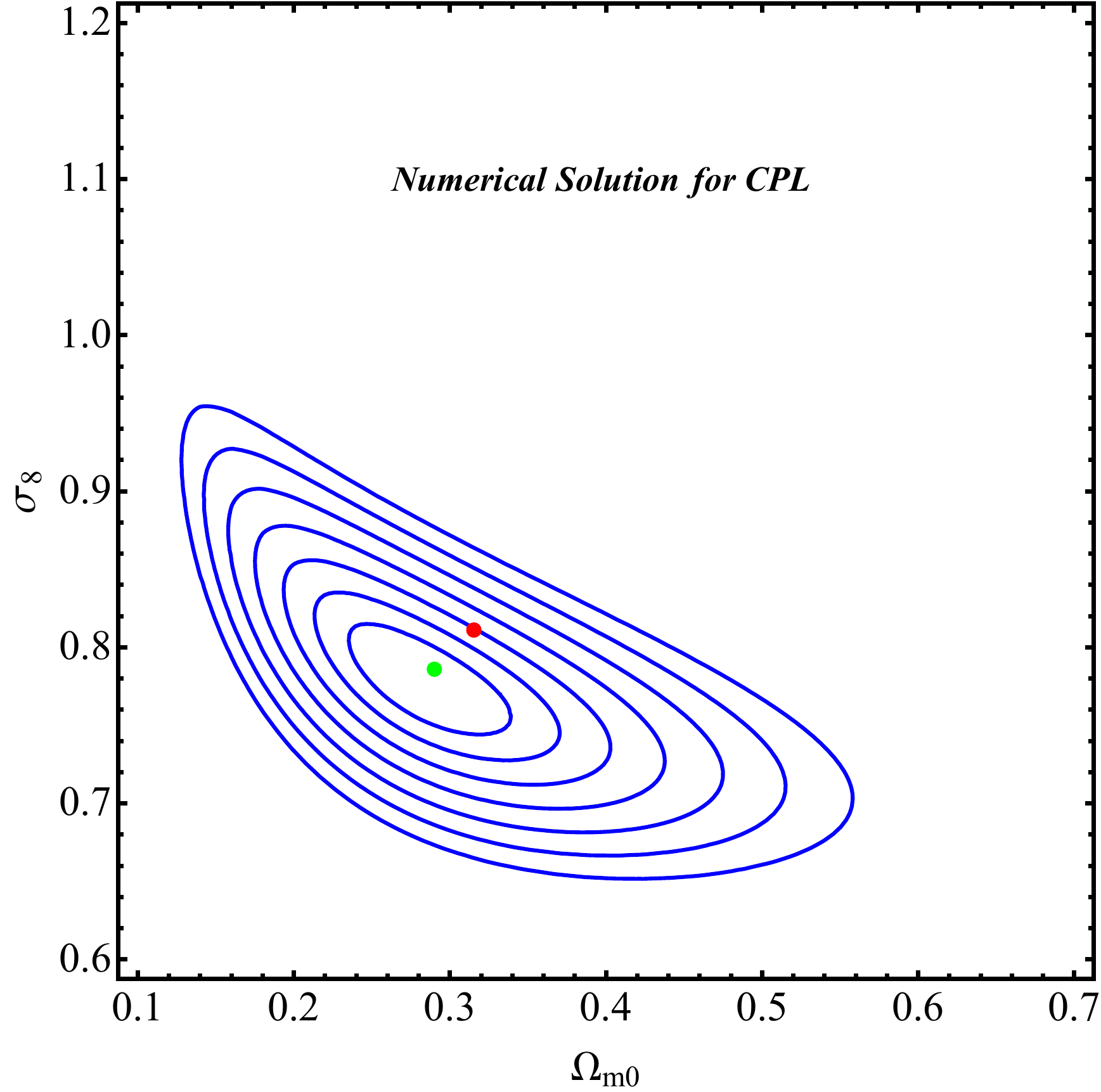}
	\includegraphics[scale=0.400]{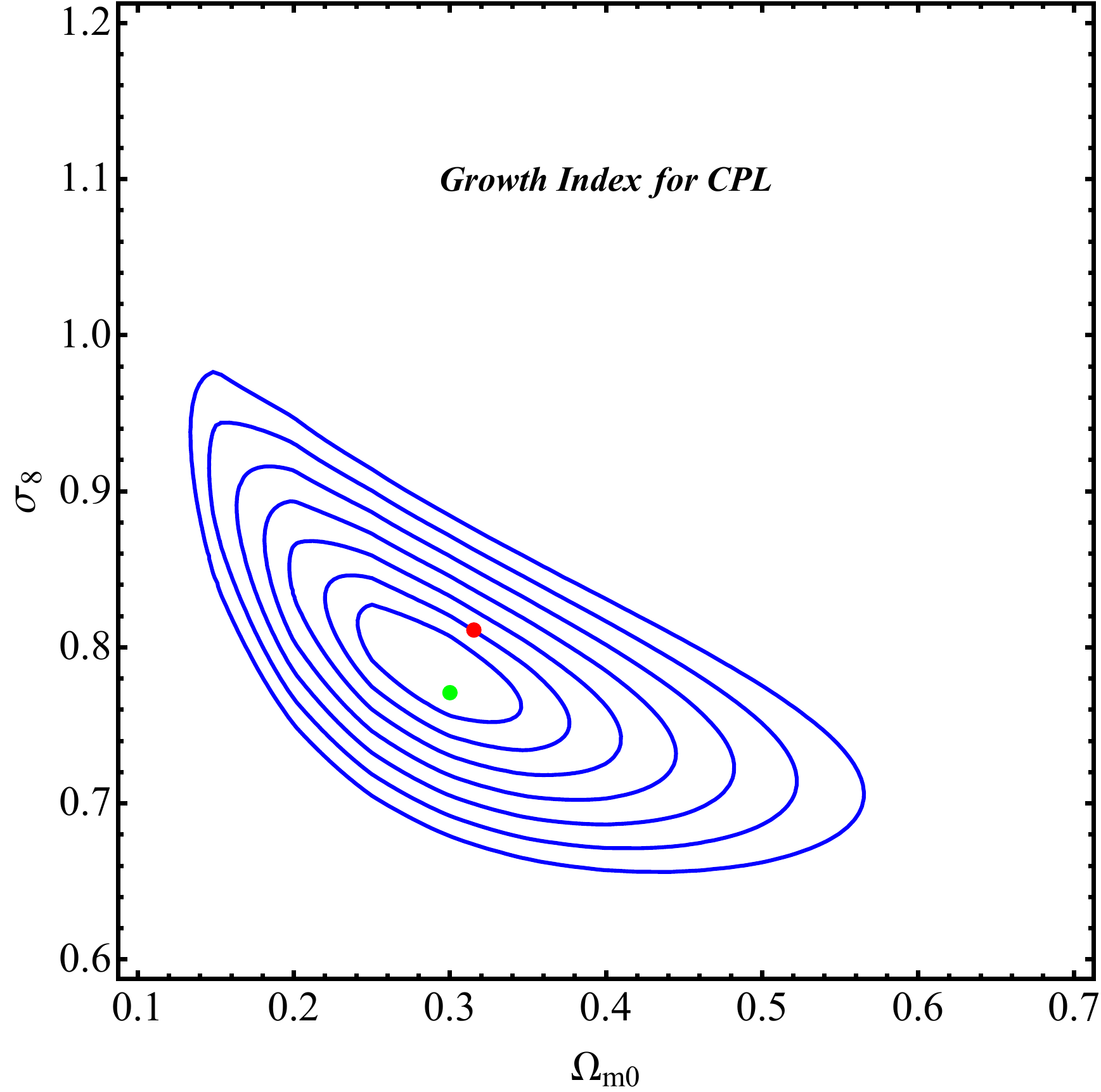}
	\includegraphics[scale=0.400]{numericalsolutionklcdm.pdf}
	\includegraphics[scale=0.400]{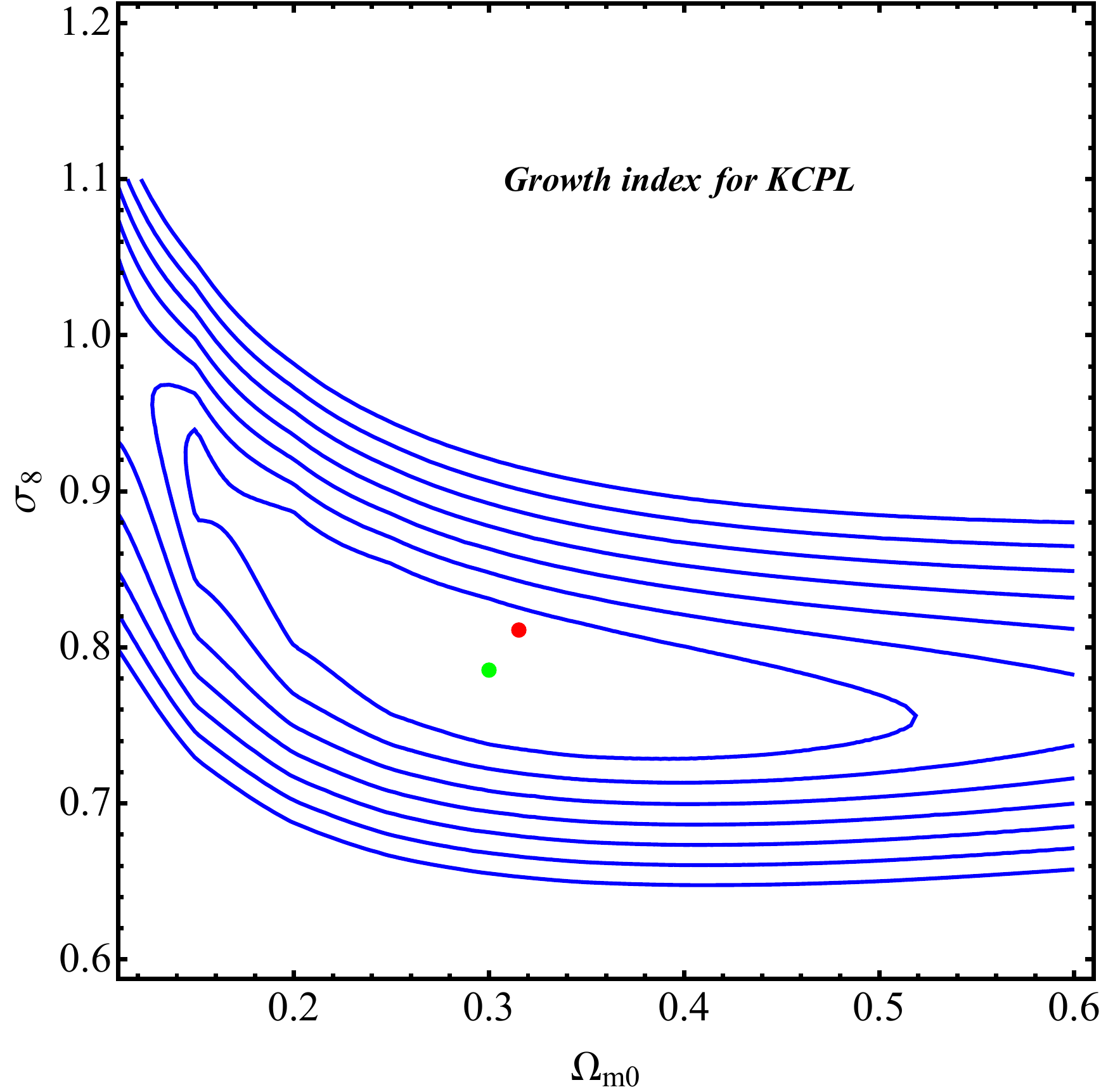}
 	\caption{In the top, we can see observational constraints on the flat $w_{0} w_{a} CDM$ model.
 	In the bottom, we see the observational constraints on the non-flat $w_{0} w_{a} CDM$ model.
	The $\Omega_{k0}$ parameter was marginalized in the range of
$-0.1<\Omega_{k0}<0.1$. We using the best fitting for $w_{0} =-0.900$, $w_{a}=-0.204$ and $\gamma_{0}=0.561$ and $\gamma_{a}=0.068$.}
 	\label{fig:epsilon_01_03}
 \end{figure*}

\begin{figure*}[htbp] 
 	\centering
 		\includegraphics[scale=0.400]{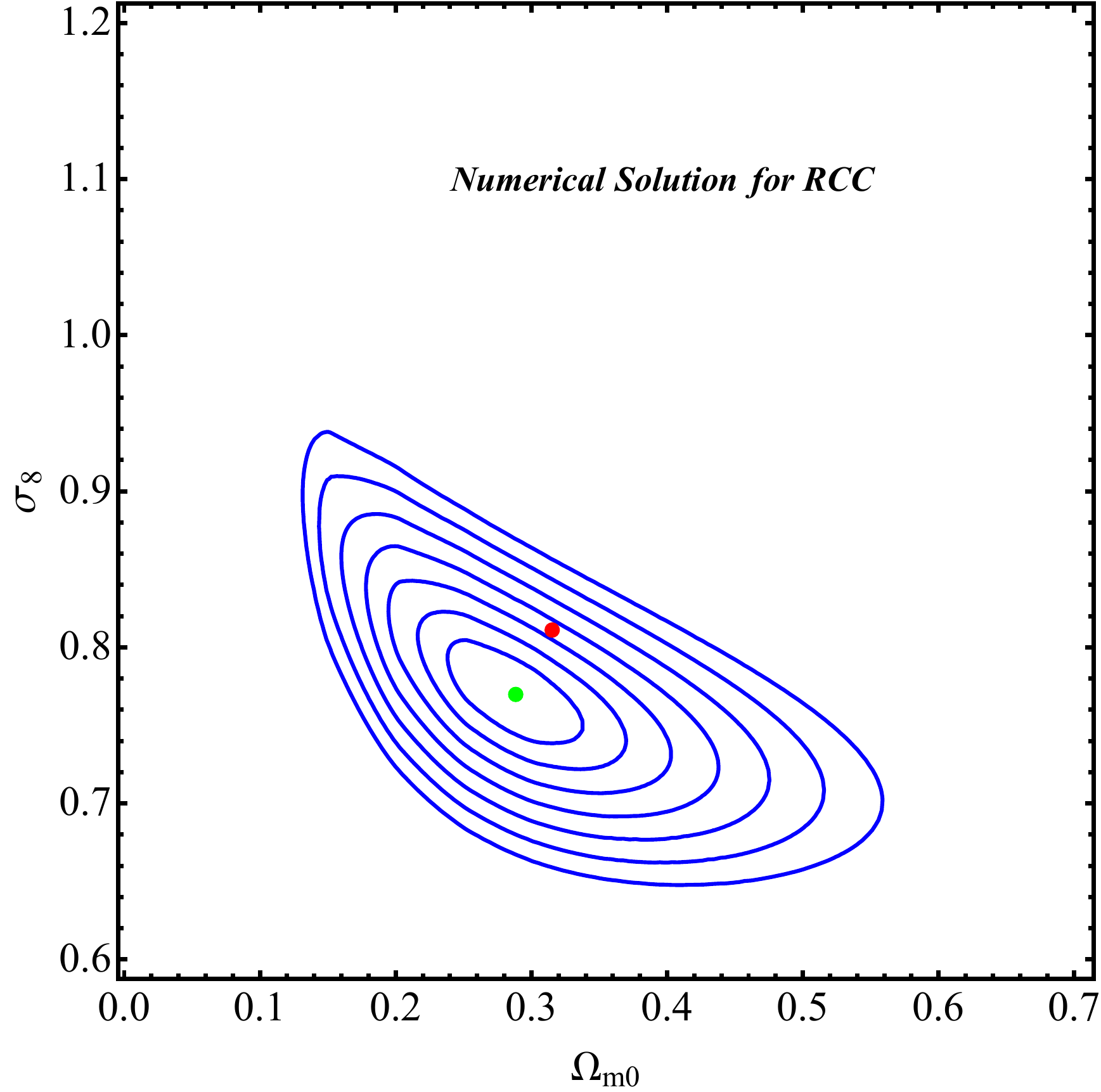}
		\includegraphics[scale=0.400]{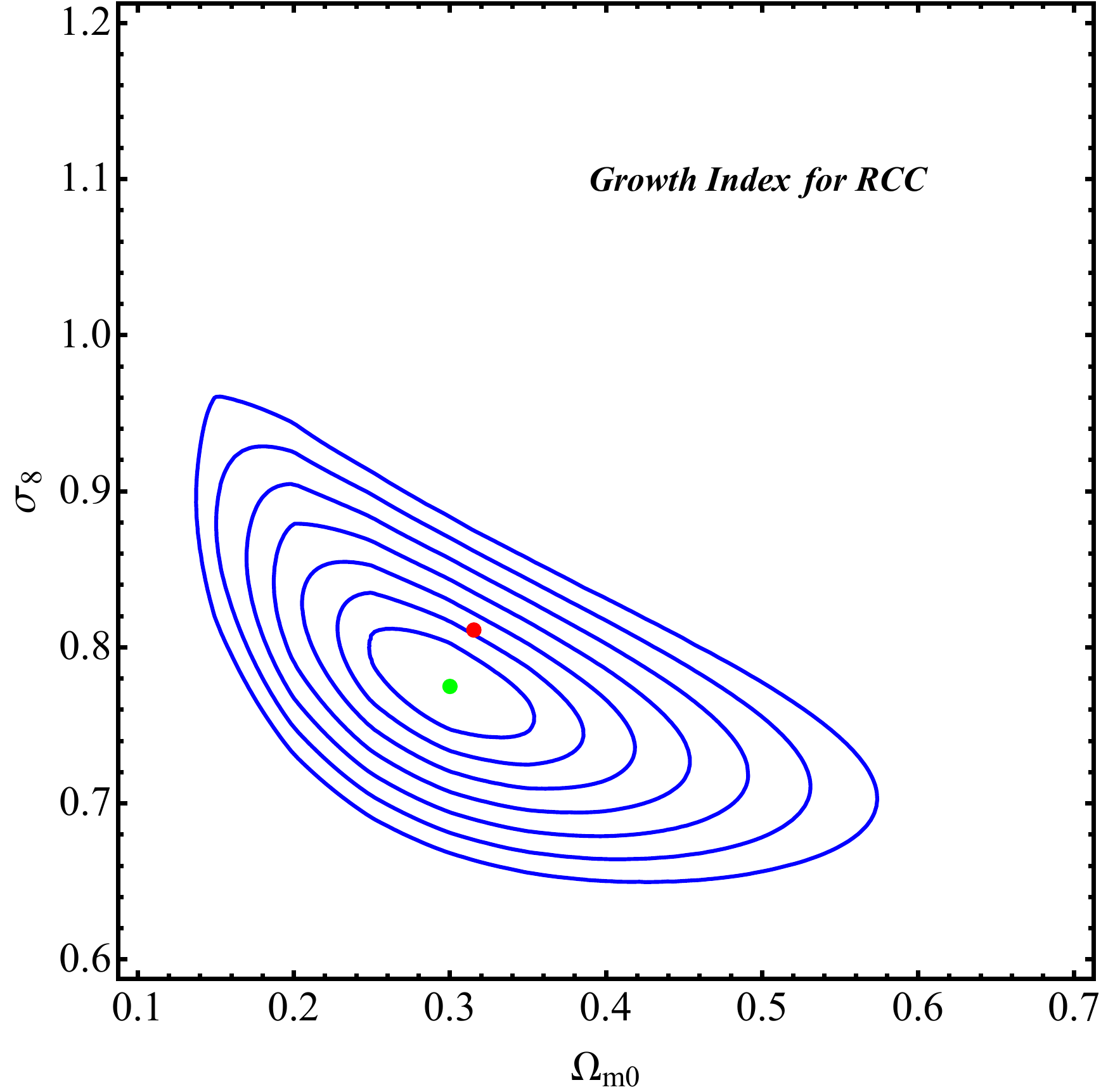}
		\includegraphics[scale=0.400]{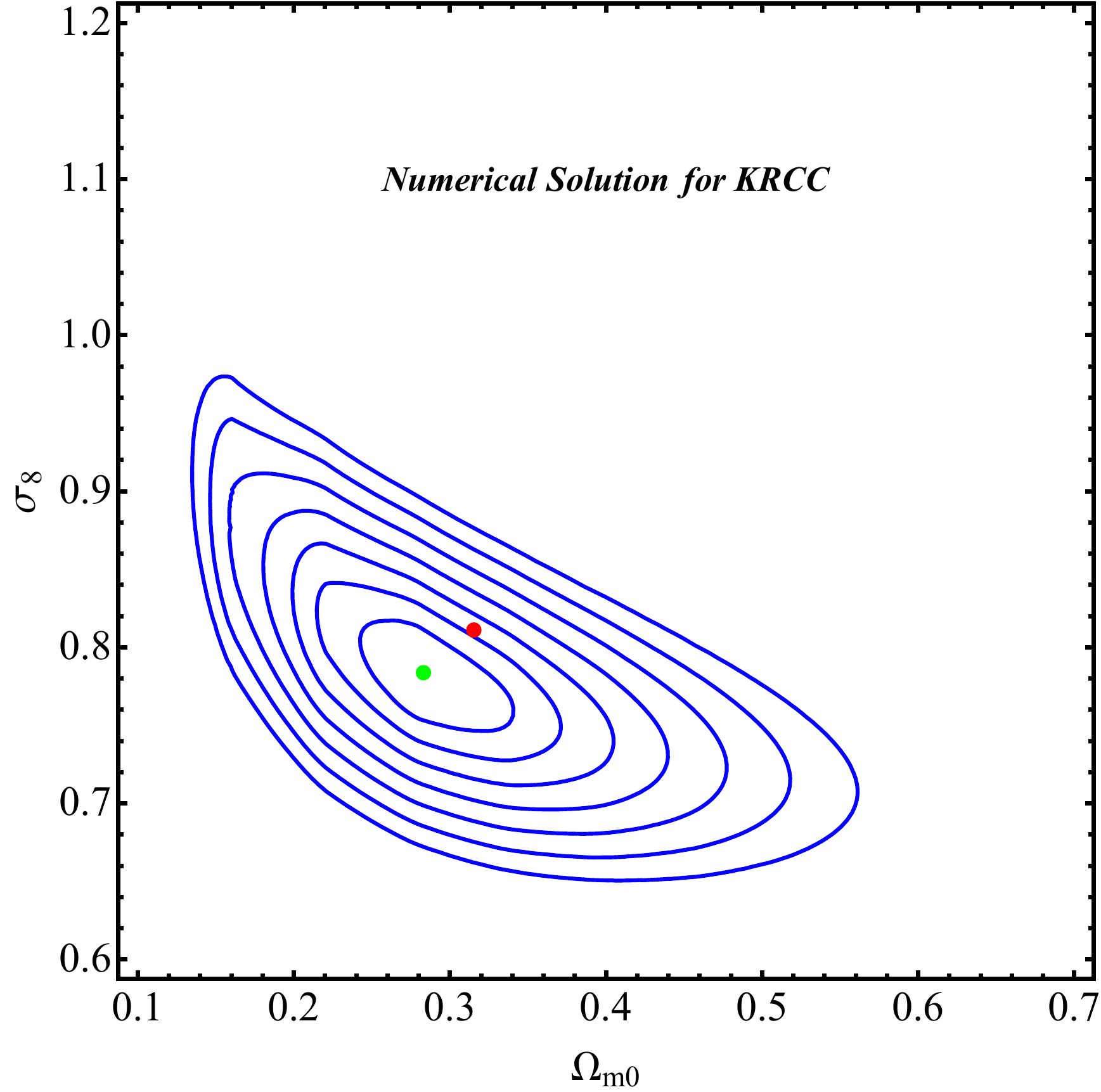}
		\includegraphics[scale=0.400]{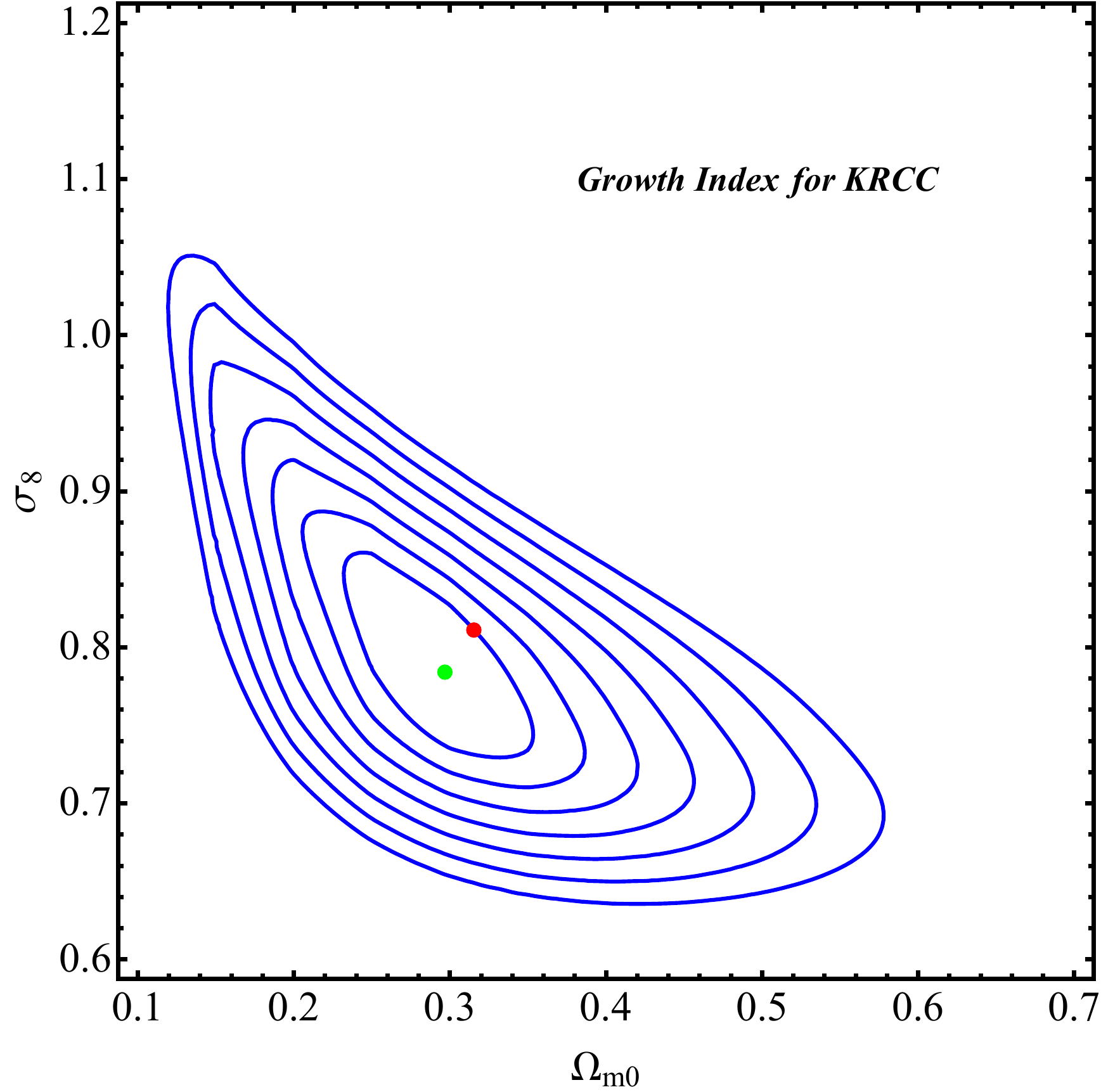}
 	\caption{In the top, we can see observational constraints on the flat $RCC$ model.
 	In the bottom, we see the observational constraints on the non-flat $RCC$ model. 
	The $\Omega_{k0}$ parameter was marginalized in the range of
$-0.1<\Omega_{k0}<0.1$.}
 	\label{fig:epsilon_01_03}
 \end{figure*}
\begin{figure*}[htbp]
\centering
\includegraphics[scale=0.300]{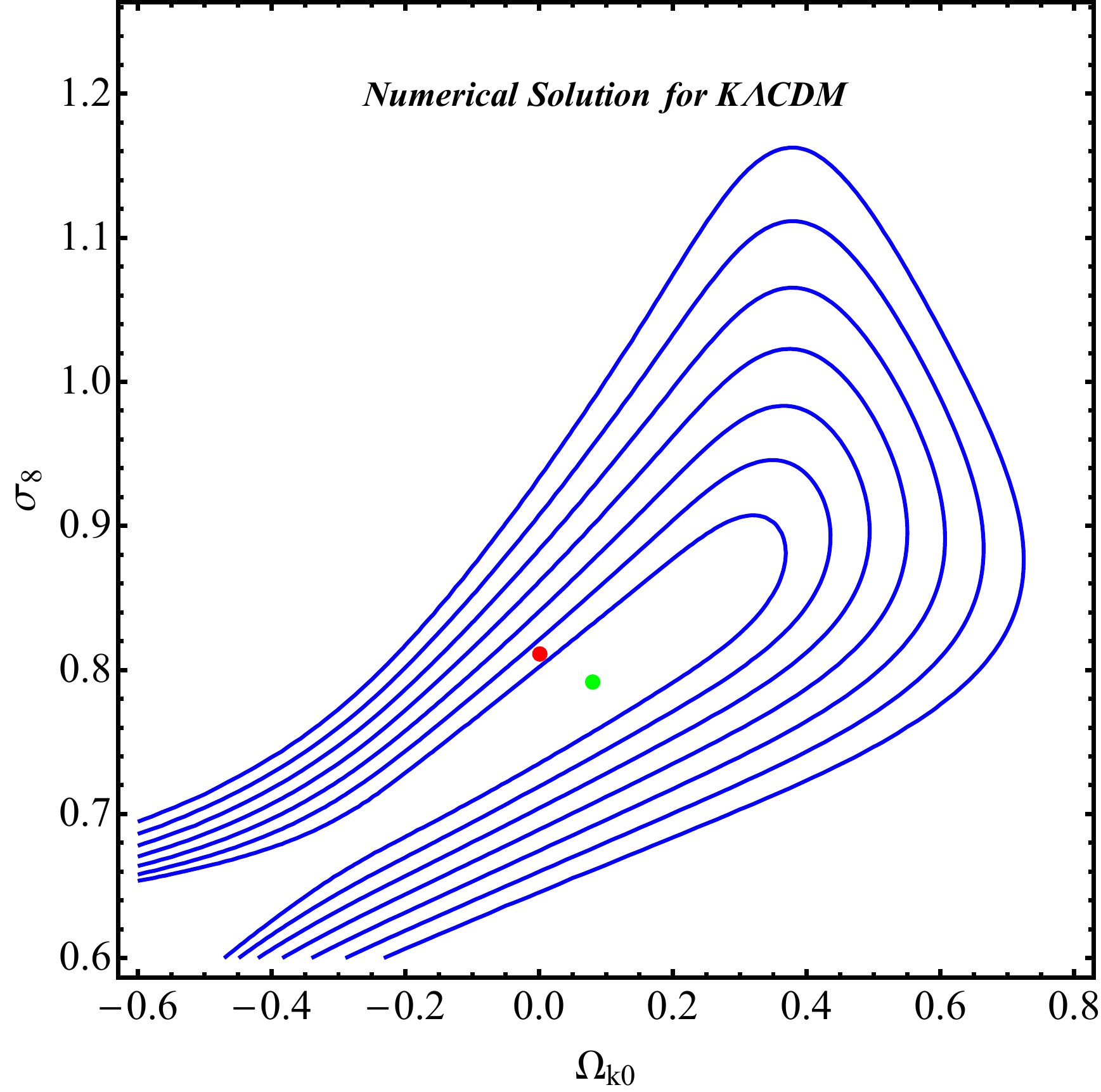}
\includegraphics[scale=0.300]{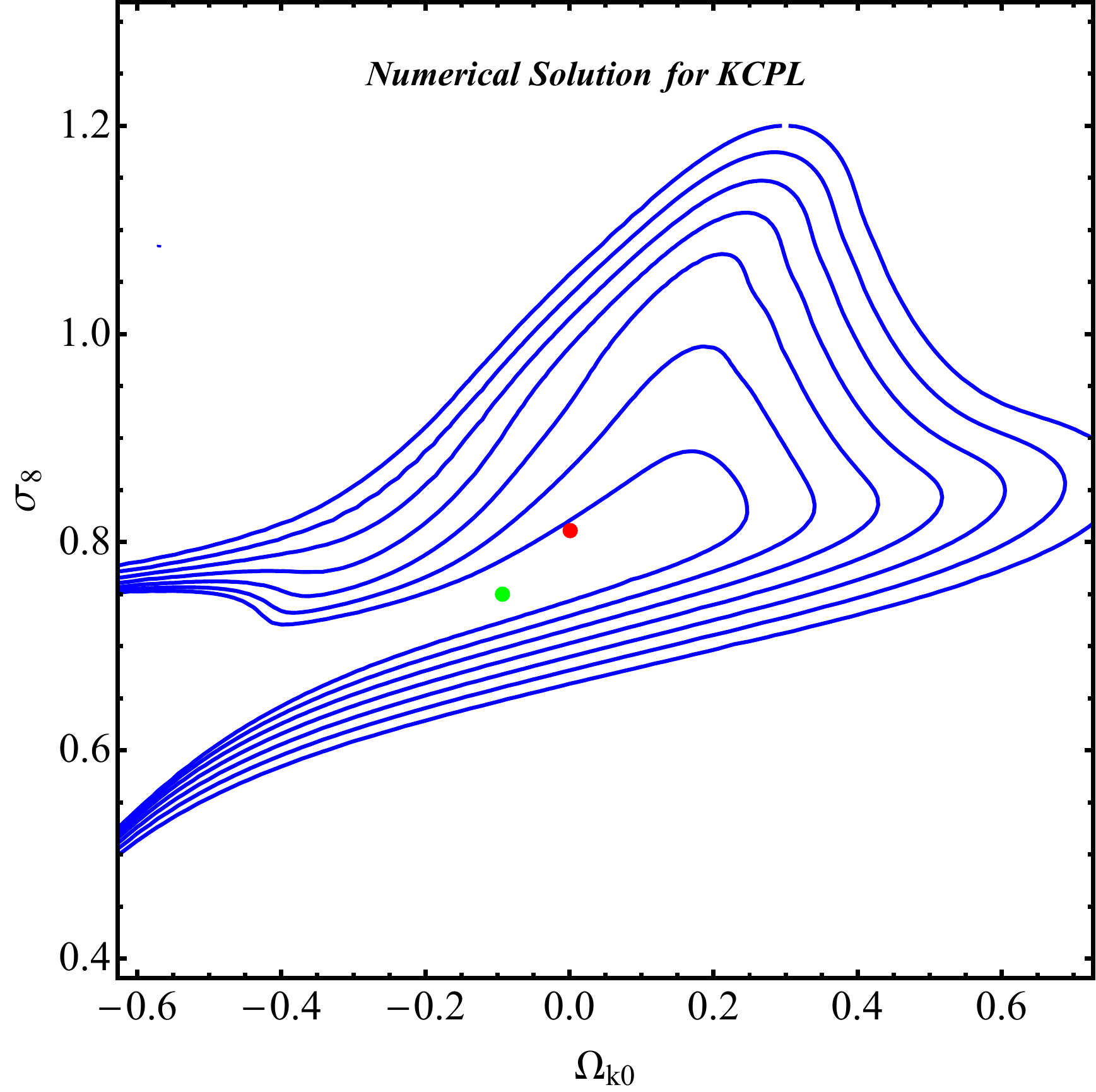}
\includegraphics[scale=0.300]{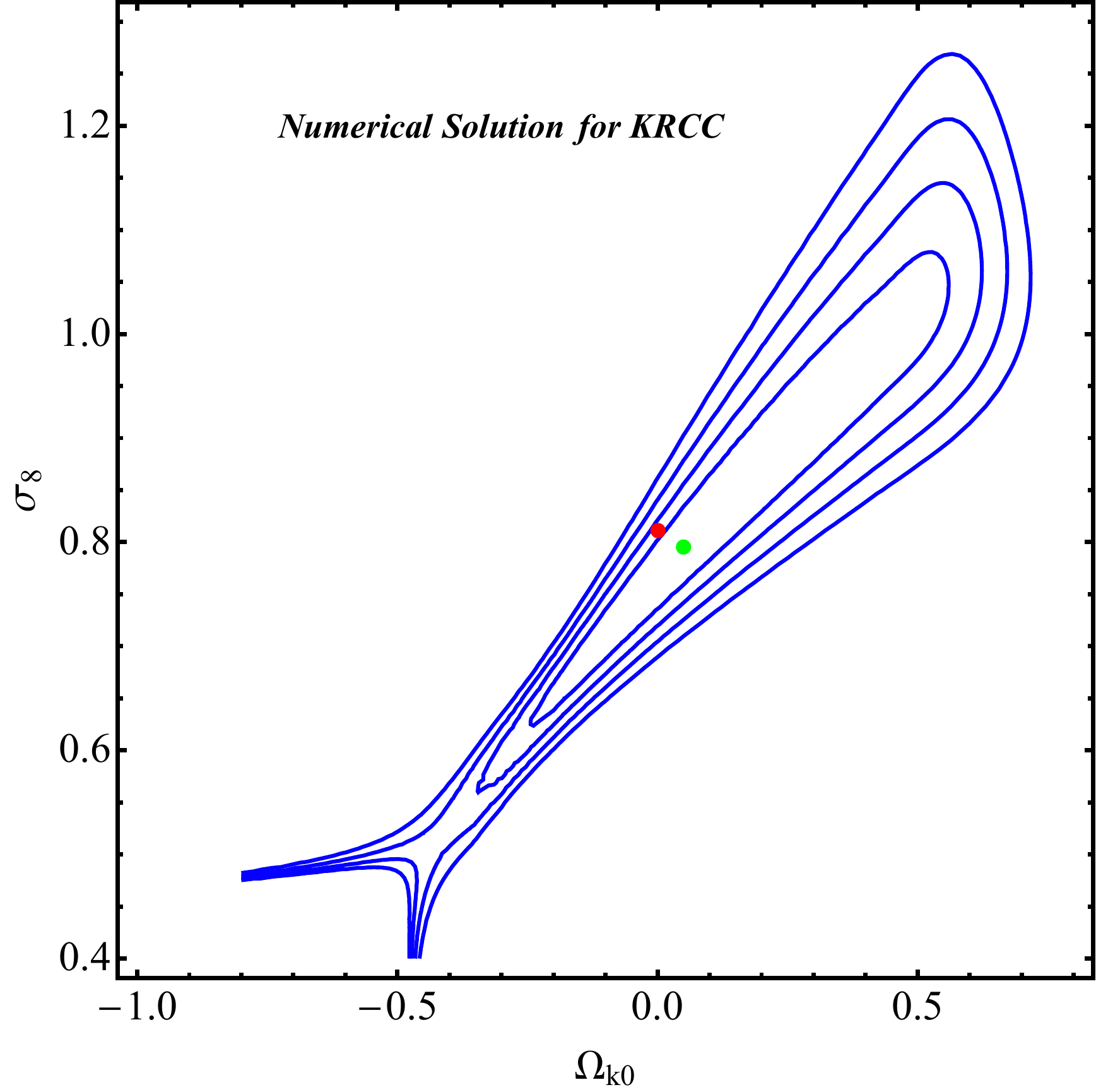}
\includegraphics[scale=0.300]{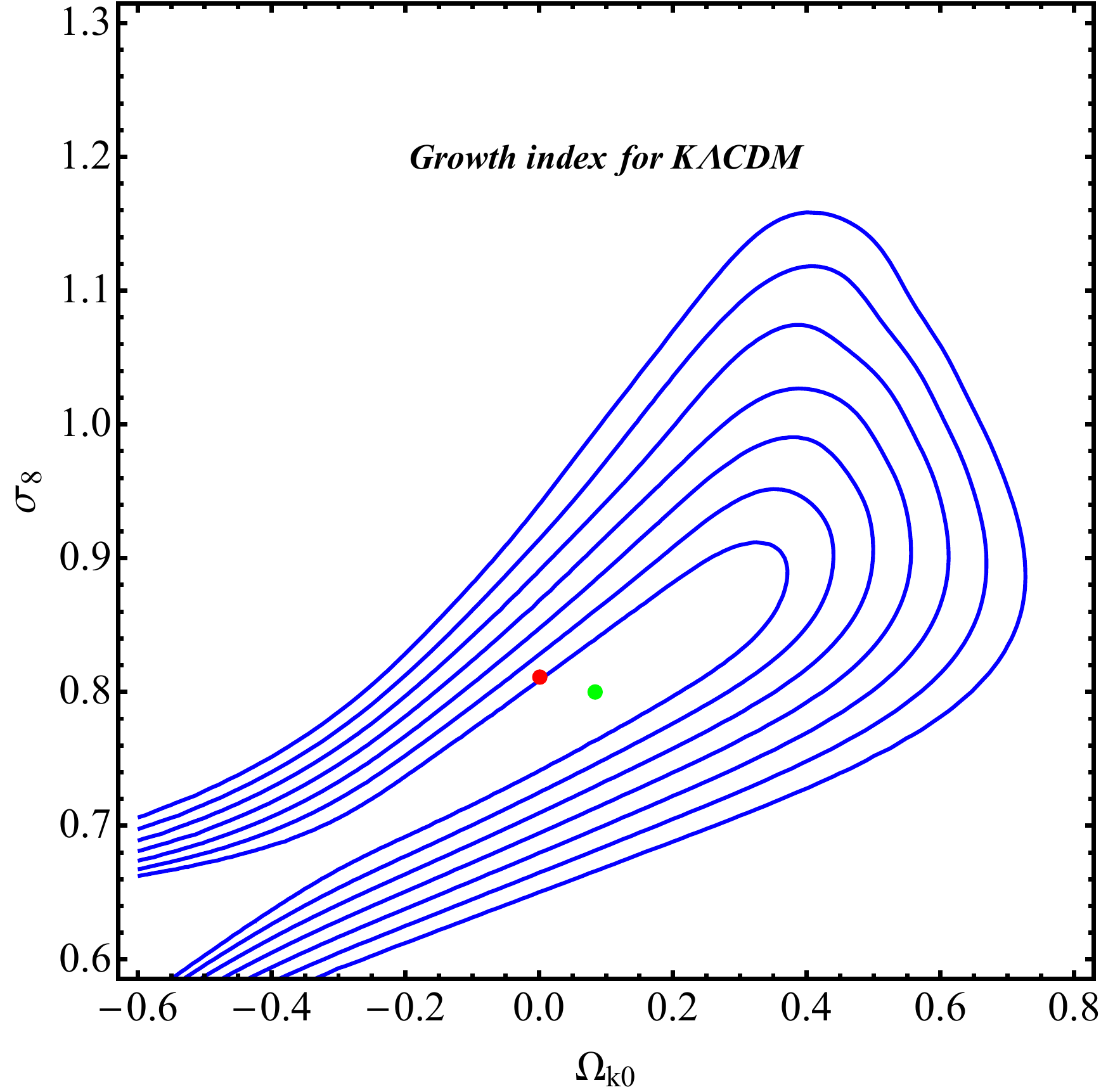}
\includegraphics[scale=0.300]{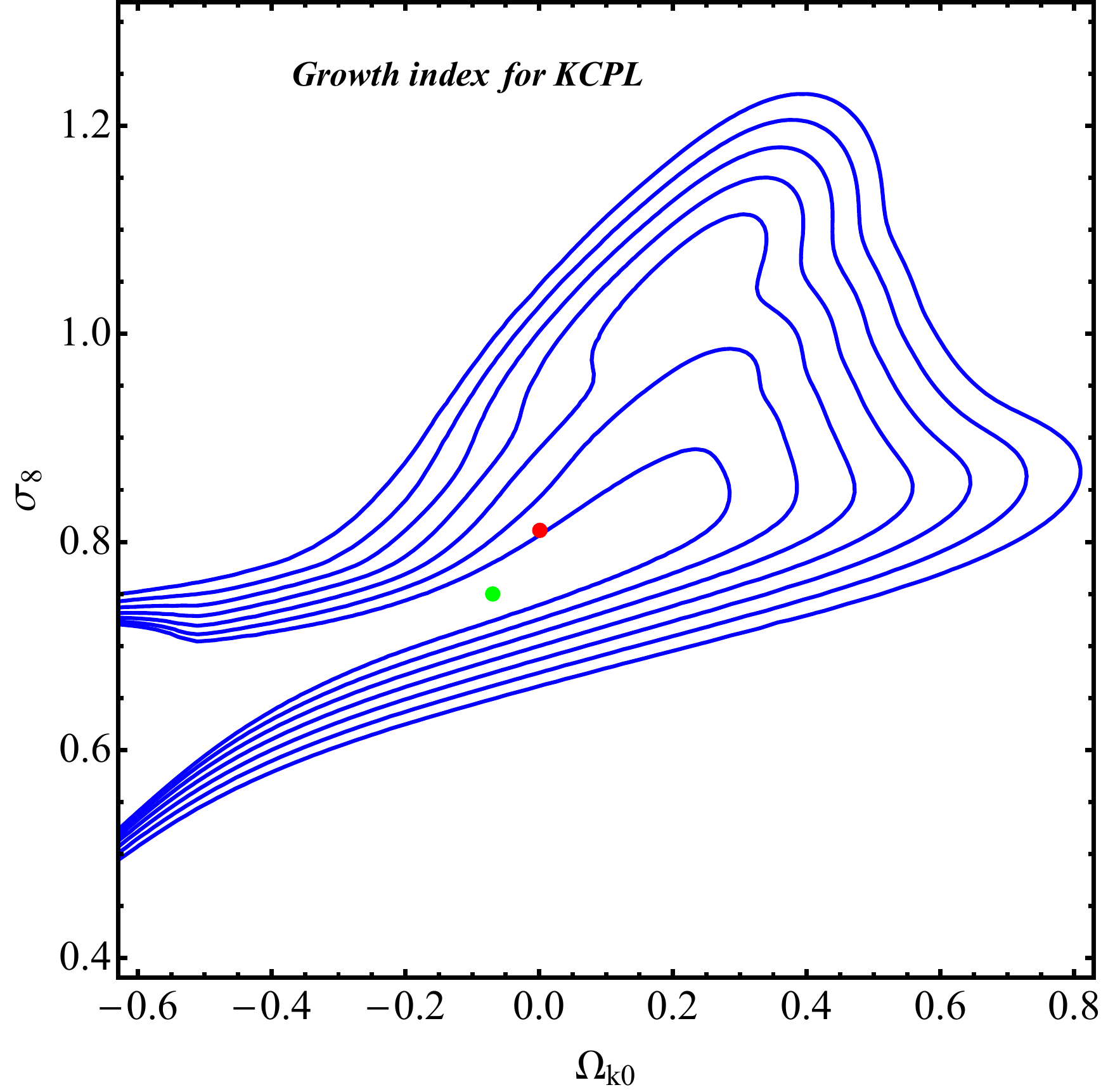}
\includegraphics[scale=0.300]{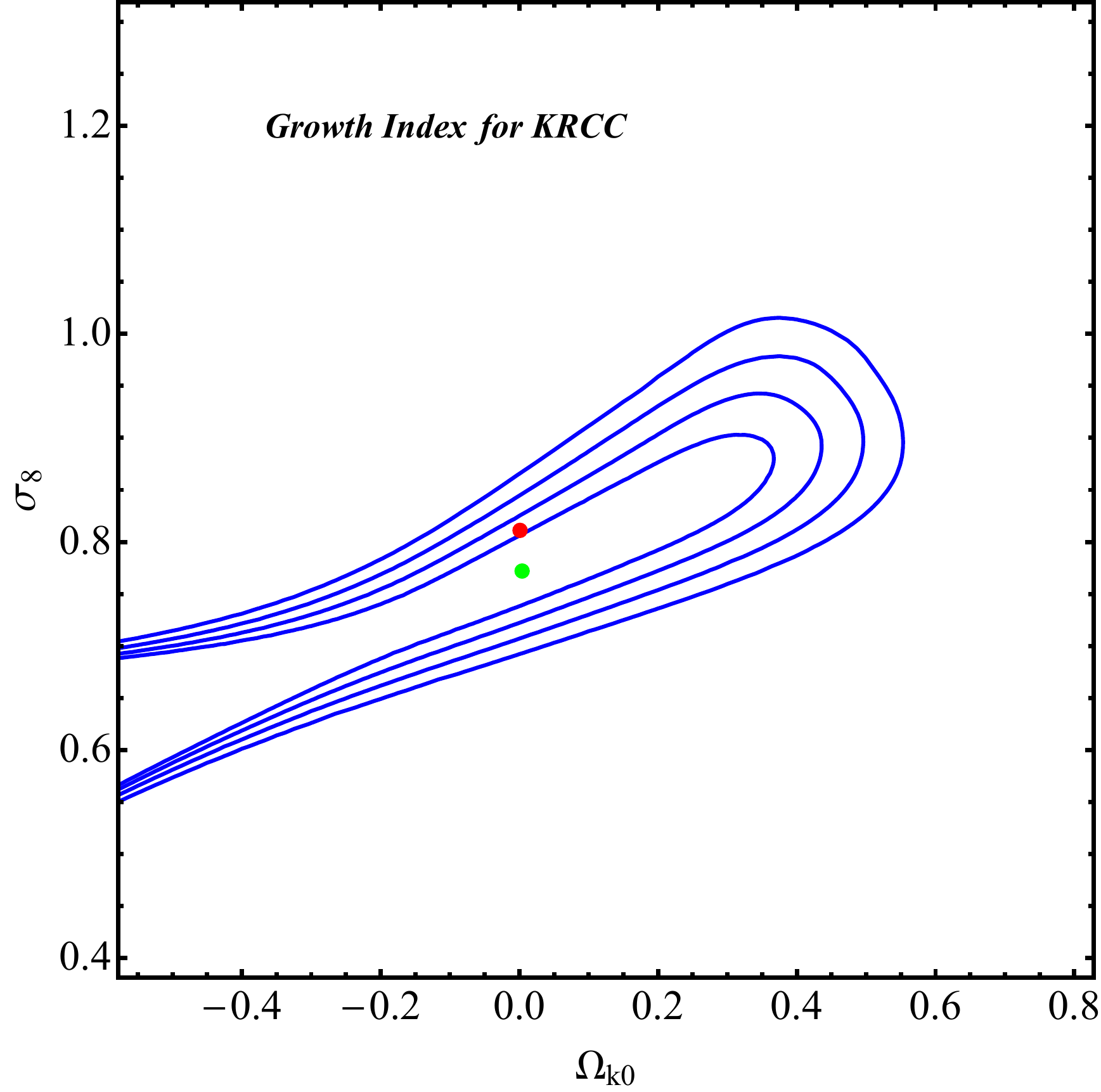}
\caption{In the top, we can see observational constraints on the models using numerical solution.
 	In the bottom, we see the observational constraints on the models using growth index.
	The $\Omega_{m0}$ parameter was marginalized in the range of $0.1<\Omega_{m0}<0.5$.}
 	\label{fig:epsilon_01_03}
 \end{figure*}

\begin{figure*}[htbp] 
 	\centering
	\includegraphics[scale=0.75000]{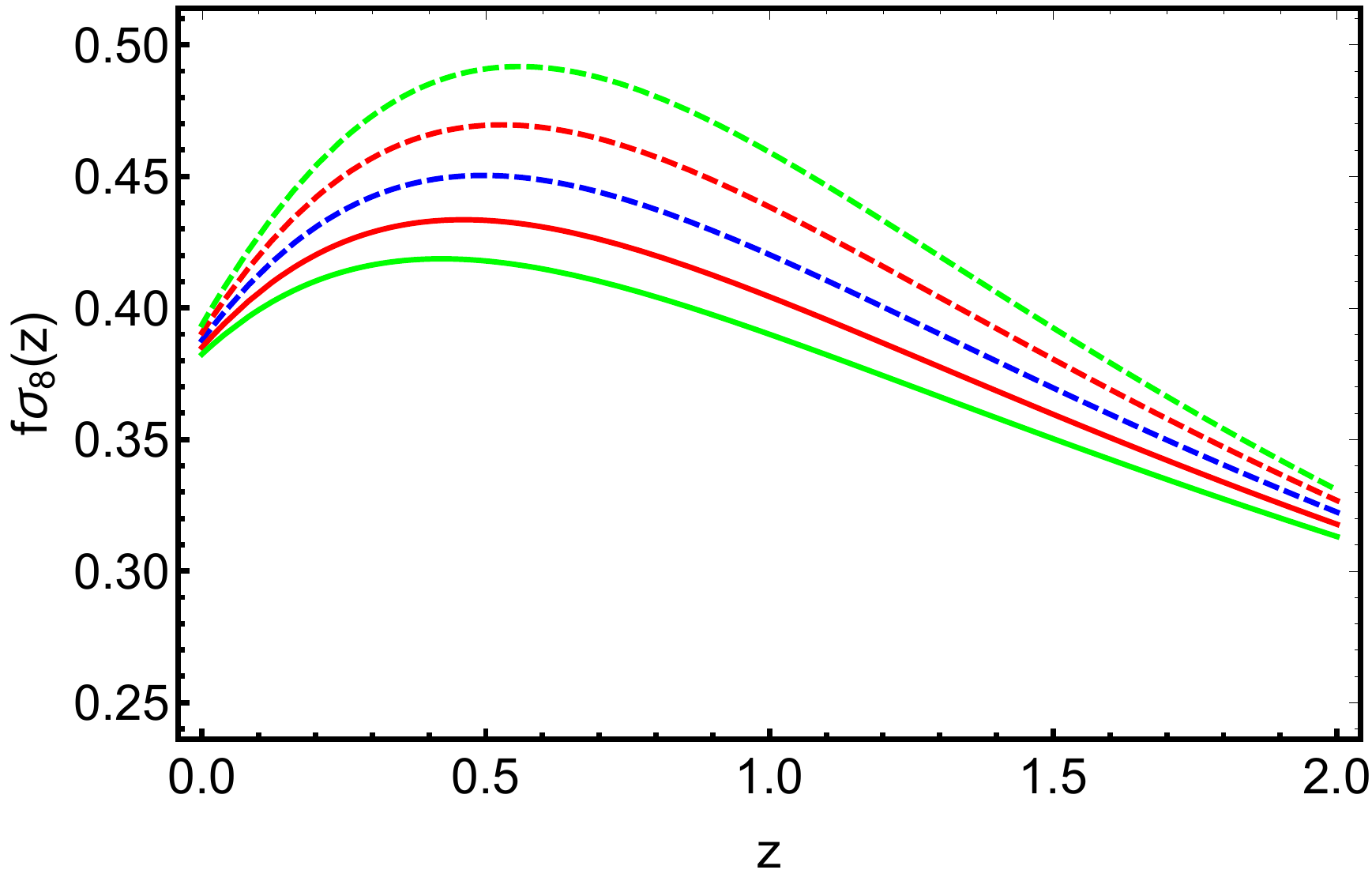}
\includegraphics[scale=0.7500]{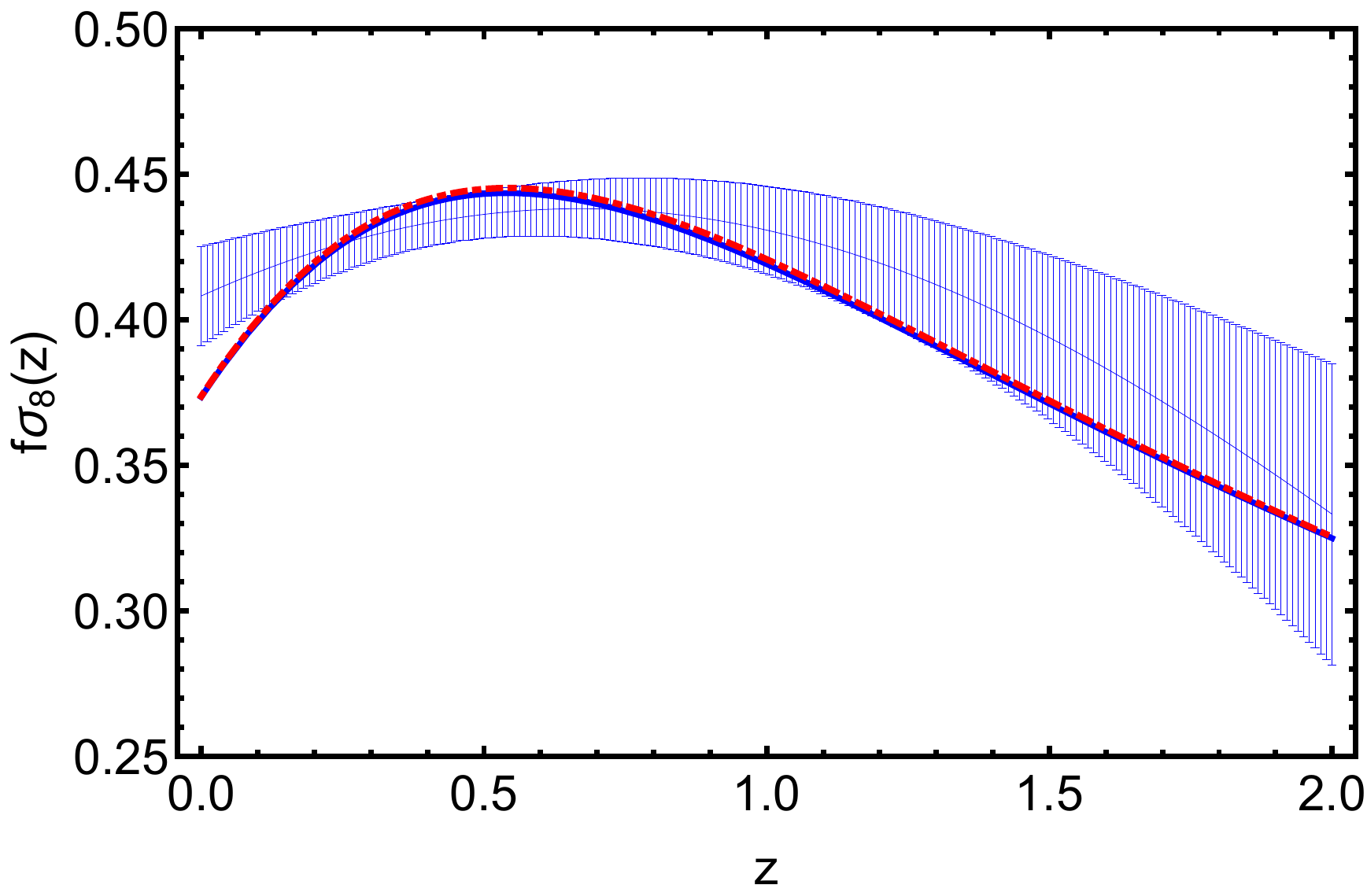}
\caption{In the top, we can see the $ f \sigma_{8} $ observable for the $ \Lambda CDM $ model for different values of the curvature parameter using numerical solutions. In the bottom, we shown the non-parametric reconstruction of the observable $ f \sigma_{8} $ and the blue curve represents the best fit for the flat case and the dashed red curve represents the best fit for the non-flat case.}
 	\label{fig:epsilon_01_03}
 \end{figure*}

\begin{table}\center
\begin{tabular}{|l|l|l|}
\hline

\multicolumn{3}{|c|}{$K\Lambda CDM$} \\
\hline
Parameter & Numerical Solution & Growth Index \\ \hline \hline
$\Omega_{m0}$ & $0.277 \pm 0.118$        & $0.277 \pm 0.165$   \\ \hline
$\Omega_{k0}$  &  $0.075 \pm 0.204$         &   $0.083 \pm 0.165$ \\ \hline
$\sigma_{8}$        &  $0.791 \pm 2.00$   &           $0.799 \pm 2.00$    \\ \hline
$\gamma_{0}$        &     &      $0.599 \pm 0.080$     \\ \hline
\hline
\multicolumn{3}{|c|}{$KCPL$} \\  
\hline
Parameter & Numerical Solution & Growth Index \\ \hline \hline
$\Omega_{m0}$        &  $0.303 \pm 0.70$     &           $0.299 \pm 0.83$    \\ \hline
$\Omega_{k0}$        &  $-0.043 \pm 0.123$    &      $-0.069 \pm 0.170$     \\ \hline
$\sigma_{8}$  &  $0.749 \pm 0.050$     &   $0.774 \pm 0.050$ \\ \hline
$w_{0}$  &  $-0.900$     &   $-0.957$ \\ \hline
$w_{a}$  &  $-0.204$     &   $-0.290$  \\ \hline
$\gamma_{0}$  &    &   $0.561 \pm 1.07$ \\ \hline
$\gamma_{a}$  &      &   $0.068 \pm 0.100$ \\ \hline
\hline
\multicolumn{3}{|c|}{$KRCC$} \\
\hline
Parameter & Numerical Solution & Growth Index \\ \hline \hline
$\Omega_{m0}$        &  $0.283 \pm 0.070$     &           $0.299 \pm 0.81$   \\ \hline
$\Omega_{k0}$        &  $0.065 \pm 0.195$    &      $0.050 \pm 0.132$     \\ \hline
$\sigma_{8}$  &  $0.795 \pm 0.204$     &   $0.770 \pm 181$ \\ \hline
$\nu$  &  $0.00001 \pm 0.00005$     &   $0.005 \pm 0.012$ \\ \hline
$\gamma_{0}$  &     &   $0.58 \pm 0.212$ \\ \hline
$\gamma_{a}$  &      &   $-0.01 \pm 0.070$ \\ \hline
\end{tabular}
\caption{Best-fitting parameters $1\sigma$ confidence intervals.}
\end{table}

\section{Results}
Our results for the $\Lambda CDM$ model are shown in figure 1. where we can see that for the flat case both the parametrization and the numerical result provide equivalent results. However, when we include the curvature parameter the parametrization provides better compatibility between the data of $ f \sigma_8 $ and the Planck data ($ 1 \sigma $). In general, for the $\Lambda CDM$ model, we can notice an equivalence between the two methods when determining observational constraints on cosmological parameters.
In figure 2 we show the results for the $ w_{0} w_{a} CDM $ model. We can see that in the flat case the correspondence is remarkable. However, when we introduce the curvature as a free parameter, the effect of a greater number of parameters is observed in the more open contours.

In figure 3 we shown the results for the $RCC$ model with and without curvature. We can see that in both cases the correspondence is remarkable. The $RCC$ model is the quite competitive when we consider all the models investigated in the present work. Our results are consistent with the results published in the literature on this model \cite{sola}, which show the advantages of having a model with dynamic dark energy and with the minimum number of free parameters.

We also investigated the effect of the curvature parameter. For this, in figure 4 we show the observational links in the $ (\Omega_{k0}, \sigma_{8}) $ plane. We note that both methods are equivalent for the three investigated models. It is interesting to mention that in the case of the $ CPL $ model, the best fit for the $\Omega_{k0}$ parameter corresponds to a closed model.

In figure 4. In the top we shown the $ f\sigma_{8} $ observable for different values of the curvature parameter. We can observe that
the highest sensitivity of the $ f\sigma_{8} $ observable is in the range: $0.5 <z <1.0$. In the figure bottom, we present the non-parametric reconstruction using Gaussian processes and the best fit curves (blue curve for $\Lambda CDM$ and red dashed curve for $K\Lambda CDM$). We can see that into current precision we cannot distinguish between flat and non-flat  
$\Lambda CDM$ models.

\section{Conclusions}
In the present work, we use two different methods to determine observational constraints on three cosmological models: the $\Lambda CDM$ model, the $\omega_ {0} \omega_{a} CDM$ model and the $RCC$ model (in all cases we include the curvature parameter). The first method is the parametrization of the growth factor and the second method consists of numerical solutions of the equation for the density contrast of matter, $\delta_{m}$. The data used are structure formation data and are given in function of the observable $ f\sigma_{8} $.

Specifically, we study the parameter spaces $ (\Omega_{m0}, \sigma_ {8}) $ and $ (\Omega_ {K0}, \sigma_ {8}) $.
We show the best fits within $ 1 \sigma $ in Table 1. We can see that the parametrization does not come into tension with the numerical results and explicitly justifying the use of the parametrized version. This verification has not been considered in the literature and in view of the large amount of research using the parametrized version, we believe that it justifies showing this correspondence directly in the derivation of observational constraints.

Additionally, we also study the power of the data $ f \sigma_{8} $ for constraints the curvature parameter. In figure 5. we show that the non-parametric reconstruction of the 
$ f \sigma_{8} $ does not allow to differentiate between a flat and non-flat universe. We can also identify that for the data $ f \sigma_{8} $ the interval in redshift $ 0.5 <z <1.00 $ is the most sensitive to the curvature parameter. In this sense, observational projects such as Euclid or LSST \footnote{Information about the project Euclid, see the web page: https://www.euclid-ec.org/ and on the LSST: https://www.lsst.org/.} will allow us to obtain a greater number of data within this redshift interval and may be fundamental to determine the curvature parameter.

\section*{Acknowledgments}
A.M.V.T appreciate the computational facilities of the UFES to develop the work.
 J. C. Fabris thanks Fundação de Amparo Pesquisa e Inovação do Espirito Santo (FAPES, project number 80598935/17)
and Conselho Nacional de Desenvolvimento Científico e Tecnológico (CNPq, grant number
304521/2015-9) for partial support.

\end{document}